\documentclass[%
 preprint,
 amsmath,amssymb,
 aps,
]{revtex4-2}
\usepackage{physics}
\usepackage{mathtools}
\usepackage{bbm,bbold}
\usepackage{hyperref}
\usepackage{ amssymb }
\usepackage{lipsum}
\usepackage[labelformat=simple]{subcaption}

\usepackage{xfrac}
\usepackage{float}
\usepackage{graphicx}
\usepackage{dcolumn}
\usepackage{bm}

\begin{document}

\preprint{APS/123-QED}

\title{Numerical Simulation of a Two-Dimensional Blume-Capel Ferromagnet in an Oscillating Magnetic Field with a Constant Bias}

\author{Celeste Mendes}
\email{17-10374@usb.ve}

\author{Gloria M. Buendía}%
 \email{buendia@usb.ve}
\affiliation{Department of Physics, Universidad Simón Bolívar, Caracas 1080, Venezuela
}

\author{Per Arne Rikvold}%
 \email{p.a.rikvold@fys.uio.no}
 \affiliation{PoreLab, NJORD Centre, Department of Physics, University of Oslo, P.O. Box 1048 Blindern, 0316 Oslo, Norway}
\affiliation{Department of Physics, Florida State University, Tallahassee, Florida 32306-4350, USA}




\date{\today}

\begin{abstract}

We perform a numerical study of the kinetic Blume-Capel (BC) model to find if it exhibits the metamagnetic anomalies previously observed in the kinetic Ising model for supercritical periods [P. Riego, P. Vavassori, A. Berger, Phys. Rev. Lett. \textbf{118}, 117202 (2017); G.M. Buendía and P.A. Rikvold, Phys. Rev. B \textbf{96}, 134306 (2017)]. We employ a heat-bath Monte Carlo (MC) algorithm on a square lattice in which spins can take values of $\pm 1, 0$, with a non-zero crystal field, subjected to a sinusoidal oscillating field in conjunction with a constant bias. 
In the ordered region, we find an equivalent hysteretic response of the order parameters with its respective conjugate fields between the kinetic and the equilibrium model. In the disordered region (supercritical periods), we observed two peaks, symmetrical with respect to zero bias, in the susceptibility and scaled variance curves, consistent with the numerical and experimental findings on the kinetic Ising model. This behavior does not have a counterpart in the equilibrium model. Furthermore, we find that the peaks occur at higher values of the bias field and become progressively smaller as the density of zeros, or the amplitude of the oscillating field, increases. Using nucleation theory, we demonstrate that these fluctuations, as in the Ising model, are not a critical phenomenon, but that they are associated with a crossover between a single-droplet (SD) and a multi-droplet (MD) magnetization switching mechanism. For strong (weak) bias, the SD (MD) mechanism dominates. We also found that the zeros concentrate on the droplets' 
surfaces, which may cause a reduced interface tension in comparison with the Ising model 
[M.~Schick and W.~H.\ Shih, Phys. Rev. B {\bf 34}, 1797 (1986)]. Our results suggest that metamagnetic anomalies are not particular to the kinetic Ising model, but rather are a general characteristic of spin kinetic models, and provide further evidence that the equivalence between dynamical phase transitions and equilibrium ones is only valid near the critical point. 

\end{abstract}

\maketitle


\newpage
\section{\label{sec:intro} Introduction}

\begin{figure}[H]
    \centering
    \includegraphics[scale=0.65]{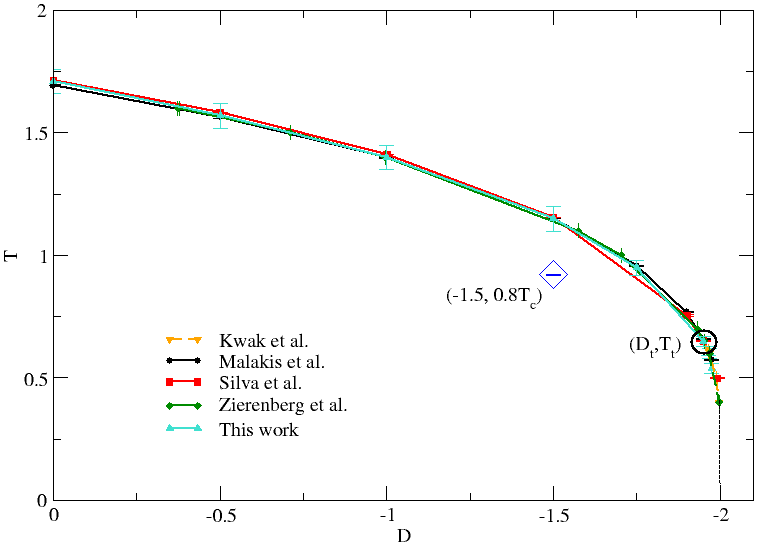}
    \caption{Temperature vs crystal-field phase diagram of the equilibrium Blume-Capel model 
    in zero magnetic field. The solid (dotted) line denotes the second-order (first-order) region, and the circled point signals the tricritical point ($D_t, T_t$) = ($1.966(1), 0.608(1)$) \cite{Kwak2015}, at which both curves join smoothly. The data correspond to the results of previous works \cite{Kwak2015, Malakis2006, Silva2006, Zierenberg2017}, as well as results from our work. The blue diamond represents the values of $D$ and $T$ used for the majority of the calculations. The critical temperature $T_c$ for $D = -1.5$ as well as for others values of $D$ can be found in Table \ref{tablapcritico}.} 
    \label{diagrama}
\end{figure}

The Blume-Capel (BC) model \cite{Blume1971, Capel1966} is a magnetic model in which spins can take values of $\pm 1, 0$. Its Hamiltonian includes a crystal field $D$, which governs the density of zeros in the system. 
The relevance and richness of this model lies in the complexities of its phase diagram \cite{Lawrie_Sarbach_1984, Per1988, Wilding1996, Selke_2010, Buendia2023}, 
some of which are displayed in Fig.~\ref{diagrama}, using data points of previous works \cite{Kwak2015, Malakis2006, Silva2006, Zierenberg2017}, as well as results from our own simulations. The system undergoes a phase transition between a paramagnetic and a ferromagnetic state. For high temperatures and greater values of $D$, the transition is continuous (second-order), while for low temperatures and sufficiently negative values of $D$, it is discontinuous (first-order) \cite{Blume1971, JGBrankov_1972, PhysRevB.15.1602, Ng1978, BALCERZAK200487, Zaim2008}. The point at which the two transition orders meet is called the tricritical point ($D_t, T_t$), which is approximately ($D_t, T_t$) = ($1.966(1), 0.608(1)$) \cite{Kwak2015}.
This complex behavior allows the BC model to describe various complex physical systems. In particular, it can be considered as a lattice-gas model, in which case it successfully describes the condensation and solidification of a simple liquid \cite{Lajzerowicz1975}, 
electrochemical adsorption \cite{RIKV88}, and the behavior of binary and ternary fluids \cite{Lajzerowiczparte2, Lajzerowiczparte3}. Other applications include superfluid mixtures of He$^3$ - He$^4$ \cite{Blume1971}, microemulsions \cite{schick1986} and the dielectric properties of nanowires \cite{BENHOURIA2018499}.  

The equilibrium and metastable properties and thermodynamic phase transitions (TPT) 
of the BC model have a well-established theory and have been extensively studied using renormalization-group techniques \cite{berker1976, Branco1997, Kaufman1981, Snowman2009}, 
transfer-matrix methods \cite{Kimel1992,Fiig1994}, 
mean-field calculations \cite{Fiig1994,Burkhardt1977, Plascak1993, Ez2004}, 
Monte Carlo simulations \cite{Jain1980, Boccara1989,YLW91,Kimel1992,Fiig1994,Silva2006, Malakis2010, Malakis2012,Buendia2023}, and exact calculations at very low temperatures \cite{CIRI96,CIRI24}. 
However, fewer studies have been conducted on its kinetic counterpart \cite{Buendia1998, Keskin2005, Keskin2007, Shi2014, Acharyya2017}. In the kinetic BC model, the system is subjected to an oscillating magnetic field of period $P$. Below its critical temperature and at a certain critical period $P_c$, the system undergoes a dynamic phase transition (DPT), associated with a competition between the period of the oscillating field and the metastable lifetime. For a slow oscillating field ($P > P_c$, disordered region), the time-dependent magnetization follows the external field, resulting in a complete reversal of the spins toward the stable state. In this case, the order parameter, the average magnetization per cycle $\langle Q \rangle$, oscillates around zero. On the other hand, for fast oscillations ($P < P_c$, ordered region), the spins do not have enough time to reverse in a single cycle, the system stays in one of its equilibrium states, 
and $\langle Q \rangle \neq 0$.

For values of $D$ corresponding to the second-order region, the kinetic BC model belongs to the same universality class as the Ising model \cite{Vatansever2018}. Therefore, it is worth asking whether other phenomena observed in the kinetic Ising model are also present in the kinetic BC model. In an experimental study on thin Co films, Riego et al.~\cite{Riego2017} found an anomalous behavior in the dynamic susceptibility and the scaled variance when a constant bias 
(which has been previously confirmed to be the conjugate field of $\langle Q \rangle$ \cite{Robb2007,Robb2008}), is added to a slow ($P \geq P_c$) oscillating field. These anomalies are characterized by two sharp peaks, symmetrical with respect to zero bias. 
This behavior has no analogue in the corresponding thermal phase transition and, given that previous studies indicated strong similarities between the kinetic and equilibrium models \cite{Sides1998, Buendia2008, Park2012, Idigoras2012, Park2013}, it was unexpected.
Further numerical studies on the kinetic Ising model also found these ``sideband'' peaks  \cite{Buendia2017} and confirmed that they are not due to a critical phenomenon or residual magnetization, but rather to a change in the magnetization-reversal mechanisms with the bias. Since then, these metamagnetic anomalies have been studied extensively with experimental \cite{PhysRevE.102.022804, PhysRevE.108.064121}, computational \cite{YUKSEL2021126172, PhysRevE.108.034125}, and mean-field techniques \cite{SHI2019120998, Bati_2017}.

A recent mean-field study on the kinetic BC model \cite{Yuksel2022}, in absence of a crystal field ($D = 0$), showed that, when a bias field is added, this model also exhibits the fluctuation peaks in the susceptibilities in the disordered region. 

The purpose of this work is to perform an exhaustive numerical study, using Monte Carlo (MC) simulations, of a two-dimensional kinetic BC ferromagnet on a square lattice of size $L \times L$, with nearest-neighbor interactions and a crystal field $D$, subjected to a sinusoidal external field of period $P$ and amplitude $H_0$, together with a constant bias $H_b$. We carry out our calculations with $D \leq 0$, the case in which the zeros are favored, while maintaining $D$ values that belong to the second-order region (see Fig.~\ref{diagrama}). We study the dependence of the system on the bias in the ordered ($P < P_c$), and disordered regions ($P > P_c$). While in the ordered region we find a similar behavior to that of the equilibrium model, in the disordered region we find symmetrical fluctuation peaks in the susceptibility curves, analogous to those observed in the experiments conducted on thin films \cite{Riego2017} and by numerical simulations 
of the $S= 1/2$ Ising model \cite{Buendia2017}. These fluctuations correspond to sudden increases in the order parameter $\langle Q \rangle$. 
Using nucleation theory for different lattice sizes, period lengths, and bias values, we demonstrate that these metamagnetic anomalies are a consequence of a change in the reversal magnetization mechanism of the system, in the regions in which the oscillating field and the bias have opposite signs. As was previously shown for the Ising model \cite{Buendia2017}, this is a stochastic-resonance phenomenon. 

The presence of these anomalies are further evidence that the well-known equivalence between the DPT and the TPT of the Ising and BC models \cite{Sides1998, Buendia2008, Park2012, Idigoras2012, Park2013, Vatansever2018, PhysRevE.97.062146, YUKSEL2020167249, PhysRevE.104.024108} is only valid in the vicinity of the critical point. Outside this region, kinetic models can display more complex behavior.

The rest of this work is organized as follows: In Sec. \ref{simulation}, we describe the Hamiltonian of the BC model and our simulation method. We also define the observables that we will measure and we calculate the critical periods for different values of $D$. In Sec. \ref{results}, we show and discuss our results. 
In Sec. \ref{disordered}, we demonstrate the appearance of sideband peaks for a supercritical period and a certain value of $D$, and we study the dependence of these bands on the different system parameters and system sizes. The results of our nucleation study for different values of $L$, $P$, $D$ and $H_b$ are found in Sec. \ref{nucleation}. Finally, we summarize our findings in Sec. \ref{conclusion}.

\section{Model and Monte Carlo simulation}
\label{simulation}

We consider a kinetic $S = 1$ Blume-Capel (BC) model on a square lattice, subjected to an external time-dependent magnetic field, along with a constant bias. 
The Hamiltonian of this system is
\begin{equation}
   \mathcal{H} = - J \sum_{\langle ij \rangle} S_i S_j - D \sum_i S^2_i- [H(t) + H_b] \sum_i S_i, 
\end{equation}
where $\langle ij \rangle$ indicates summation over nearest neighbors, and the other two summations run over all sites. $J > 0$ is the ferromagnetic exchange interaction constant and $S_i = \pm 1, 0$. $D$ is the single-ion crystal field, which distinguishes between $S_i = 0$ and $S_i = \pm 1$. As $|D|$ increases ($D \leq 0$), the number of zeros increases. $H_b$ is a constant bias field and $H(t)$ is a periodically oscillating field with period $P$ and amplitude $H_0$ given by
\begin{equation}
    H(t) = H_0 \cos{\left(\frac{2\pi t}{P} \right)}\quad .
    \label{h(t)}
\end{equation}

Monte Carlo (MC) simulations according to a heat-bath algorithm were performed on a square lattice of linear dimension $L$ with periodic boundary conditions, such that $N = L \times L$ denotes the total number of spins. 
A spin $S_i$ is chosen at random and the probability $ W (S_i \rightarrow S'_i)$ that the spin $S_i$ changes to $S'_i$ 
(which may or may not be equal to $S_i$) is 
\begin{equation}
    W(S_i \rightarrow S'_i) \propto e^{-\beta \Delta E(S_i \rightarrow S'_i)}\quad . 
\end{equation}
Here, $\beta = \sfrac{1}{k_B T}$, with $k_B$ the Boltzmann constant and $T$ the temperature, and $\Delta E(S_i \rightarrow S'_i)$ is the difference in energy associated with changing the spin $S_i$ to $S'_i$. 
All other spins act as a heat bath or reservoir and by definition remain constant. 

The time unit is one Monte Carlo step per site (MCSS), during which, on average, each site is visited once. $D$, $H_0$, $H_b$ and $T$ are given in units of $J$ (i.e., $J = k_B = 1$), and $P$ in units of MCSS.

We calculate the following observables. 

The average magnetization during the $k$th cycle of the field,
\begin{equation}
    Q_k = \frac{1}{P} \int_{(k-1)P}^{kP} m(t) dt\quad ,
\end{equation}

where $m(t)$ is the time-dependent magnetization per site,
\begin{equation}
    m(t) = \frac{1}{N} \sum_i S_i(t)\quad .
\end{equation}

The order parameter $\langle Q \rangle$ is the period-averaged magnetization over $f$ cycles,
\begin{equation}
    \langle Q \rangle = \frac{1}{f} \sum_{k=1}^{f} Q_k\quad .
\end{equation}

The fluctuations of the order parameter are given by the scaled variance,
\begin{equation}
    \chi^Q = L^2 [\langle Q^2 \rangle - \langle |Q| \rangle^2] \quad .
\end{equation}

Additionally, we calculate the dynamic susceptibility of $\langle Q \rangle$ 
with respect to the bias, 
\begin{equation}
    \chi^b = \frac{d \langle Q \rangle}{d H_b}\quad ,
\end{equation}
by numerical differentiation. 

The model exhibits a dynamic phase transition at a certain period that we will denote as $P_c$, which separates the ordered region ($P < P_c$), in which spins fail to make a complete reversal due to the rapid field oscillation and $\langle Q \rangle \neq 0$, from the disordered region $(P > P_c)$ in which the magnetization has enough time to follow the oscillating field and $\langle Q \rangle = 0$. 

The effect of the constant bias field $H_b$ is to add an asymmetry to the transition, such that $H_b$ favors the inversion of the spins toward the state in which they have the same sign as $H_b$. Near $P_c$, the bias is the conjugate field of $Q$ \cite{Robb2007}. 

The measurements were obtained by averaging over 800 cycles, following a 200-cycle stabilization. Each cycle has $P \times 800$ MCSS.

To capitalize on previous results from the kinetic BC model \cite{Vatansever2018}, we performed our calculations at $T = 0.8T_c$, where $T_c$ is the critical temperature of the model in equilibrium, and $H_0 = 0.2$ (except in Fig.~\ref{distintosH}, where $H_0$ varies from 0.4 to 1.5). $P$ varies between $P = 105$ and $P = 20000$, and the lattice size from $L = 32$ to $L = 1024$. Results with fixed size were performed with $L = 128$.

We performed the simulations with $D$ between $D = -1.75$ and 0. For these values of $D$, the number of zeros is significant and, according to Fig.~\ref{diagrama}, correspond to the second-order phase-transition region, which has been shown to belong to the same universality class as the equilibrium Ising model \cite{Vatansever2018}. Most calculations were performed with $D = -1.5$. The critical temperatures depend on $D$, and they are known with precision for the values used in this work. In particular, we used those calculated by Malakis et al. \cite{Malakis2010}, which are reported in Table \ref{tablapcritico}.

Table \ref{tablapcritico} also includes the values of $P_c$ calculated with our simulation. To our knowledge, there are no previous works that calculate the critical periods for the kinetic BC model subjected to a sinusoidal field for these different values of $D$. 

\begin{table}[h!]
    \centering
    \caption{Critical periods $P_c$ for the BC kinetic model subjected to a sinusoidal field for different values of crystal field $D$ at $T = 0.8 T_c$. The critical temperatures $T_c$ correspond to those calculated by Malakis et al. \cite{Malakis2010}}
    \begin{tabular}{|c|c|c|}
          \hline
         $D$ & $P_c$ (MCSS) & $T_c$\\
         \hline
         0 & 870 $\pm$ 10 & 1.693(3)\\
         \hline
         $-0.5$ & 720 $\pm$ 10 & 1.564(3)\\
         \hline
         $-1.0$ & 460 $\pm$ 10 & 1.398(2)\\
         \hline
        $-1.5$ & 205 $\pm$ 5 & 1.151(1)\\ 
         \hline
         $-1.75$ & 115 $\pm$ 5 & 0.958(1)\\
         \hline
    \end{tabular}
    \label{tablapcritico}
\end{table}

\section{NUMERICAL RESULTS AND ANALYSIS}
\label{results}

In the subcritical region, $T < T_c$ and $P < P_c$, the kinetic BC model 
in an oscillating field shows hysteretic behavior and a sharp 
susceptibility peak at $P_c$ \cite{Vatansever2018}, 
similar to the corresponding $S=1/2$ Ising model \cite{Robb2007,Robb2008}. 
In what follows, we discuss only the supercritical region, $P > P_c$, where the kinetic Ising model 
shows novel sideband phenomena that are not present in its equilibrium version 
\cite{Riego2017,Buendia2017}. Our goal is to quantitatively investigate by MC simulations  whether the same is the case in the kinetic BC model. 

\subsection{The kinetic Blume-Capel model in the supercritical region, $P > P_c$}
\label{disordered}

\begin{figure}[H]
     \centering
     \begin{subfigure}[b]{0.45\textwidth}
         \centering
         \includegraphics[scale=0.428]{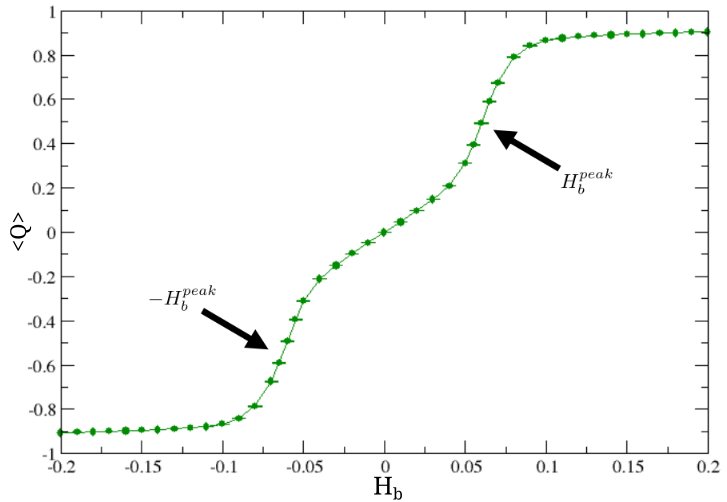}
    \caption{}
    \label{Qpeaks}
     \end{subfigure}
     \hfill
     \begin{subfigure}[b]{0.45\textwidth}
         \centering
         \includegraphics[scale=0.45]{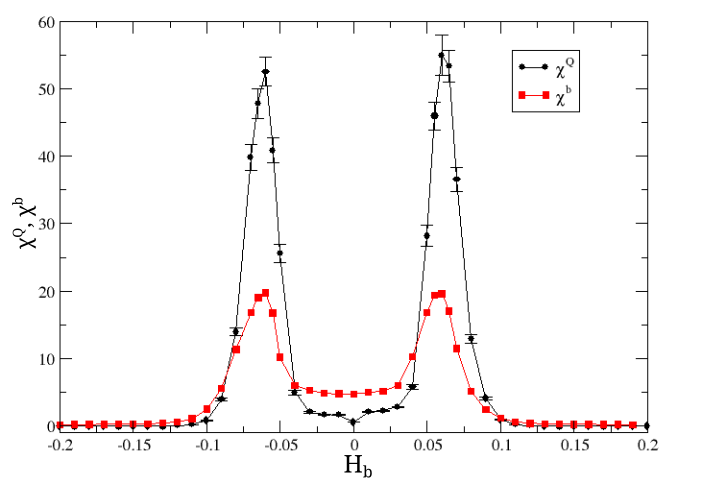}
    \caption{}
    \label{bands}
     \end{subfigure}
     \caption{Dependence of the observables of the kinetic BC model on the bias for $D = -1.5$, $L = 128$, $H_0 = 0.2$ and $P = 820 \approx 4P_c$. (a) Order parameter $\langle Q \rangle$ vs $H_b$. (b) Scaled variance $\chi^Q$ and dynamic susceptibility $\chi^b$ vs $H_b$. Sideband peaks similar to those found in the kinetic Ising model for $P > P_c$ are observed. The arrows in (a) indicate the values where the peaks occur.}
        \label{peaks}
\end{figure}

Here, we concentrate on the {\it super}critical region, 
$T < T_c$ and $P > P_c$, with the system driven by a sinusoidally oscillating magnetic 
field of period $P$ and a static bias field. 
Figure~\ref{peaks} shows the results of our simulations for $D = -1.5$ and $P = 820 \approx 4 P_c$. For this value of $D$, the kinetic BC model displays the metamagnetic anomalies or ``sideband" peaks previously observed for the kinetic Ising model \cite{Riego2017, Buendia2017}.

The dependence of the order parameter $\langle Q \rangle$ on $H_b$ is shown in 
Fig.~\ref{Qpeaks}. $\langle Q \rangle$ is an antisymmetric function with respect to $H_b$. Two regions stand out, at which the curve of $\langle Q \rangle$ increases sharply. These two regions are separated by a region of lower slope around $H_b = 0$.
This behavior is consistent with the results for the scaled variance $\chi^Q$ and the dynamic susceptibility $\chi^b$, as seen in Fig.~\ref{bands}. The two peaks, located at $|H^{peak}_b|$, correspond to the sharply increasing regions of the $\langle Q \rangle$ curve, and the flat bottom between them corresponds to the linear region of lower slope.  For these values of 
$P$ and $D$, the peaks occur at a position of $|H^{peak}_b| = 0.063 \pm 0.002$. The regions in which $\chi^Q$ and $\chi^b$ decay rapidly to zero correspond to the saturation regions in 
Fig.~\ref{Qpeaks} for large $|H_b|$.
The results of Fig.~\ref{peaks} are very similar to those obtained by previous experimental \cite{Riego2017} and computational \cite{Buendia2017} works on the kinetic Ising model. However, the two models seem to behave differently in the region around $H_b = 0$, where the susceptibilities reach a local minimum. In the Ising model (see Ref. \cite{Buendia2017}, Fig.~1(b)), $\chi^Q$ and $\chi^b$ seem to have the same values in this region, whereas for the BC model, $\chi^b$ displays a larger value than $\chi^Q$.  

\begin{figure}[H]
\centering
\hspace{-30pt}
\begin{subfigure}{0.5\columnwidth}
    \includegraphics[width=1.1\columnwidth]{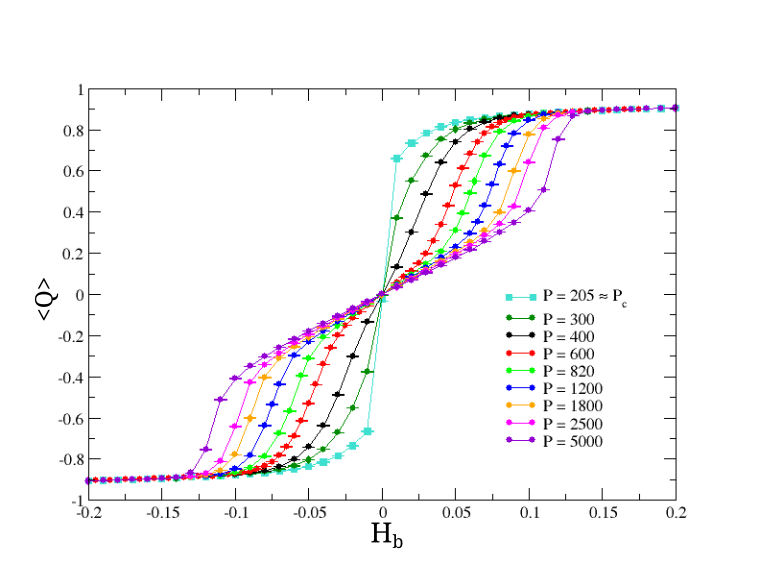}
    \caption{}
    \label{QdistintosP}
\end{subfigure}
\vspace{-7pt}
\begin{subfigure}{0.5\columnwidth}
    \includegraphics[width=1.1\columnwidth]{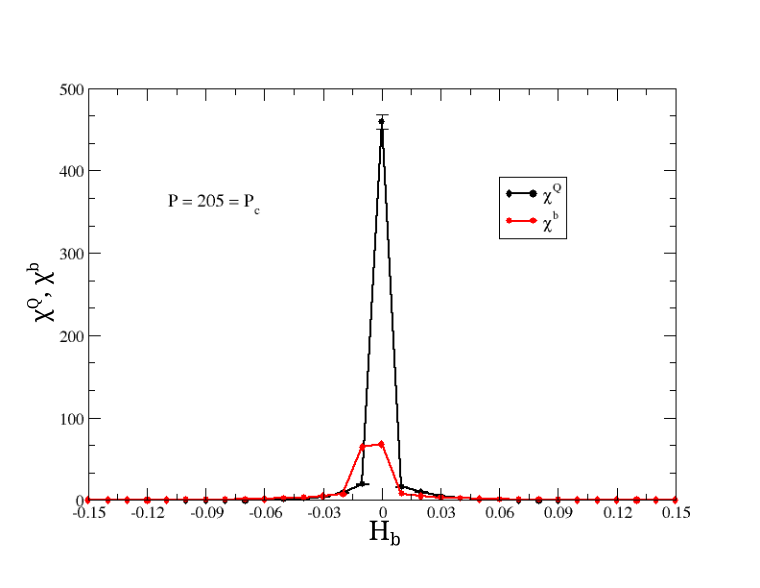}
    \caption{}
    \label{pcritico}
\end{subfigure}
\begin{subfigure}{0.5\columnwidth}
    \hspace{-30pt}
    \includegraphics[width=1.1\columnwidth]{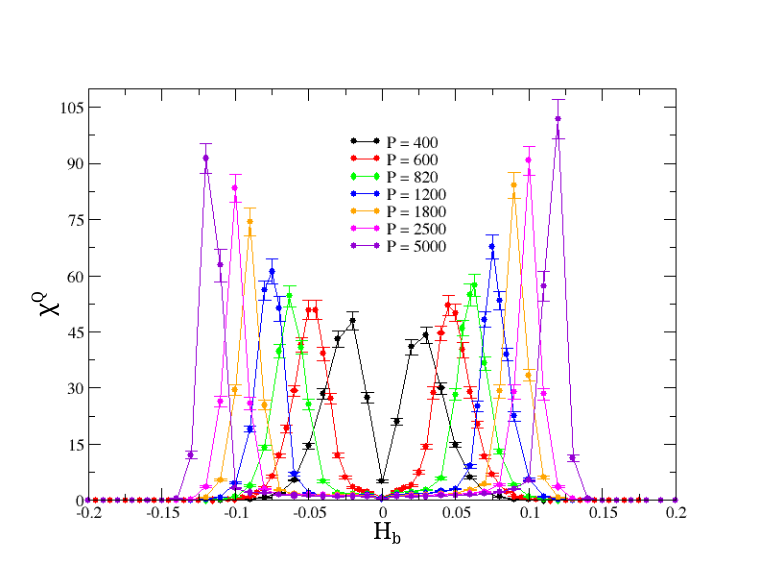}
    \caption{}
    \label{chiQdistintosP}
\end{subfigure}
\hspace{-30pt}
\begin{subfigure}{0.5\columnwidth}
    \includegraphics[width=1.1\columnwidth]{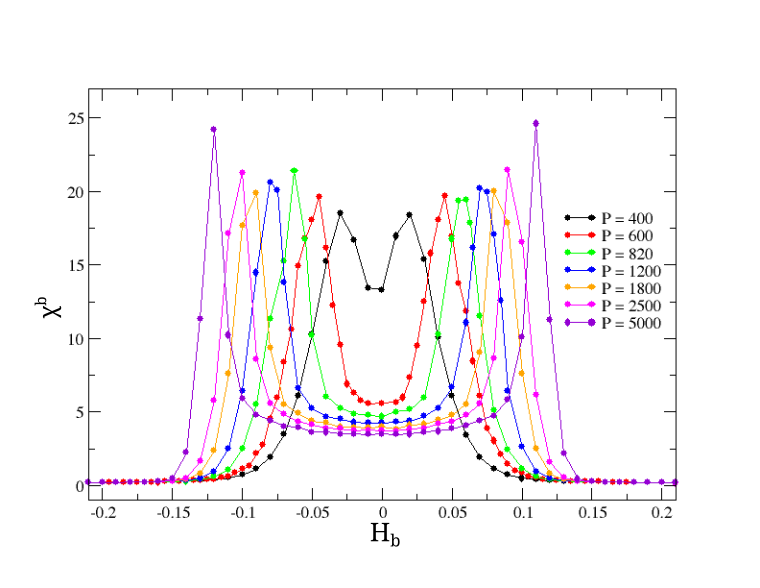}
    \caption{}
    \label{chibdistintosP}
\end{subfigure}
\caption{Dependence of the sidebands on the period $P \ge P_c$, for $D = -1.5$, $L = 128$, and $H_0 = 0.2$. (a) $\langle Q \rangle$ vs $H_b$. 
(b) $\chi^Q$ and $\chi^b$ vs $H_b$ for $P = P_c$, showing a single central peak. (c) $\chi^Q$ and (d) $\chi^b$ vs $H_b$ for different periods, in the region $P > P_c$. Two symmetric peaks are observed, whose height and separation increase with $P$.}
\label{biasdiffP}
\end{figure}

In Fig.~\ref{biasdiffP}, we present the dependence of the sidebands on the period $P$ ($P \geq P_c$) of the oscillating external field. $\langle Q \rangle$ depends strongly on $P$, as reflected in Fig.~\ref{QdistintosP}. For $P \approx P_c$, $\langle Q \rangle$ exhibits a sudden increase at $H_b = 0$. This results in a narrow peak centered at $H_b = 0$ for the susceptibilities, as seen in Fig.~\ref{pcritico}.
As $P$ increases, the transition around $H_b = 0$ smooths out. For $P > P_c$, two symmetric points are obtained with respect to $H_b = 0$ in which $\langle Q \rangle$ increases rapidly, which translates into two symmetric peaks in the susceptibilities, as shown in Figs. \ref{chiQdistintosP} and \ref{chibdistintosP}. Contrary to what happens at $P = P_c$, where these curves exhibit a peak at $H_b = 0$, for longer periods, they exhibit a minimum at $H_b = 0$.
Additionally, the separation of the peaks increases with $P$. The longer the period, the more time the spins have to reverse toward the stable state, so the asymmetry added by $H_b$ becomes effectively weaker, and a larger bias is needed to produce the sidebands. These results are in agreement with those found by previous studies on the kinetic Ising model \cite{Buendia2017, Riego2018}.\\

\begin{figure}[H]
\centering
\begin{subfigure}{\columnwidth}
    \centering
    \includegraphics[width=0.55\columnwidth]{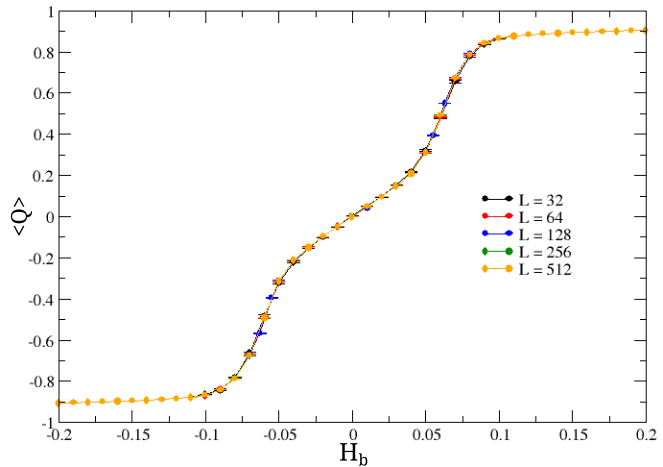}
    \caption{}
    \label{QdistintosL}
\end{subfigure}

\hspace{-35pt}
\begin{subfigure}{0.5\columnwidth}
    \centering
    \includegraphics[width=1.1\columnwidth]{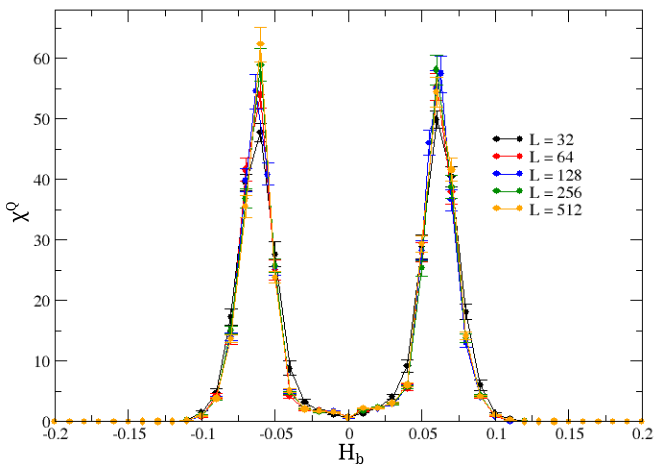}
    \caption{}
    \label{chiQdistintosL}
\end{subfigure}
\hspace{20pt}
\begin{subfigure}{0.5\columnwidth}
    \centering
    \includegraphics[width=1.1\columnwidth]{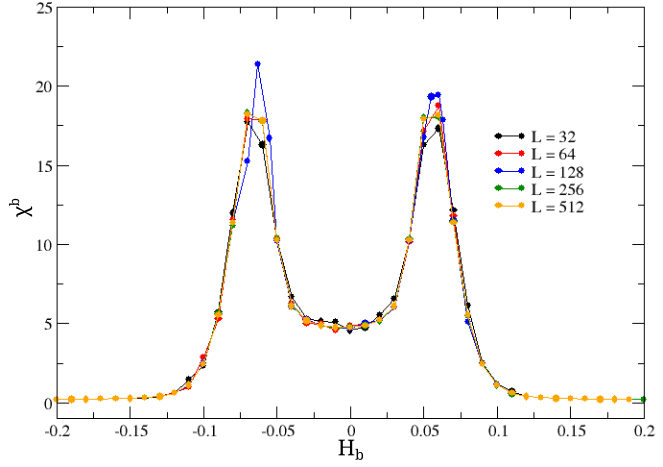}
    \caption{}
    \label{chibdistintosL}
\end{subfigure}
     \caption{Finite-size effects on the sidebands with $D = -1.5$ and $P = 820 \approx 4 P_c$ . (a) $\langle Q \rangle$ vs $H_b$. (b) $\chi^Q$ and (c) $\chi^b$ vs $H_b$. 
     In both (b) and (c), the peaks saturate as $L$ increases, indicating that they 
     are not caused by a 
     phase transition.}
        \label{distintosL}
\end{figure}

The dependence of the sidebands on $L$ for $D = -1.5$ and $P \approx 4 P_c$ is displayed in Fig.~\ref{distintosL}.
For the order parameter $\langle Q \rangle$, the curves in Fig.~\ref{QdistintosL} show no appreciable differences at this scale. 
Similarly, the position and height of the peaks of $\chi^Q$ and $\chi^b$ are not 
significantly affected by the system size, except for the two smallest values 
of $L$, as shown in Figs.~\ref{chiQdistintosL} and \ref{chibdistintosL}, respectively. 
The fact that the peak heights saturate as $L$ increases, indicates that these sidebands 
in the kinetic BC model are not the result of a phase transition. This has previously been demonstrated for the kinetic Ising model \cite{Riego2017, Buendia2017}.
The dependence of the peak positions on $L$ and $P$ is studied in more detail in 
Sec.~\ref{nucleation}, using arguments based on nucleation theory.

\begin{figure}[H]
\centering
\begin{subfigure}{\columnwidth}
    \centering
    \includegraphics[width=0.55\columnwidth]{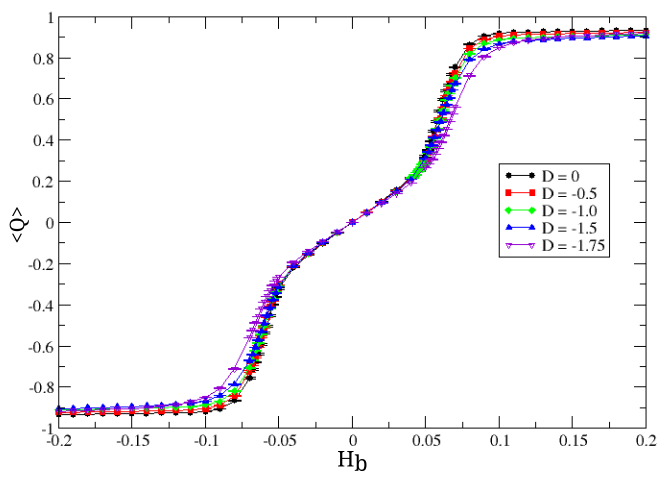}
    \caption{}
    \label{QdistintosD}
\end{subfigure}

\hspace{-50pt}
\begin{subfigure}{0.5\columnwidth}
    \centering
    \includegraphics[width=1.1\columnwidth]{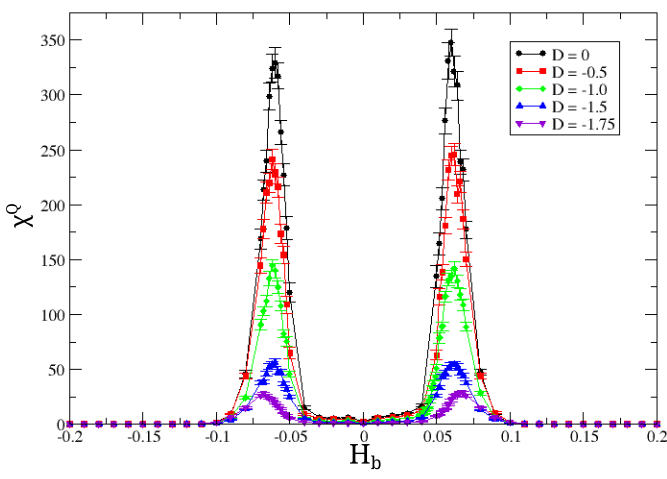}
    \caption{}
    \label{chiQdistintosD}
\end{subfigure}
\hspace{20pt}
\begin{subfigure}{0.5\columnwidth}
    \centering
    \includegraphics[width=1.2\columnwidth]{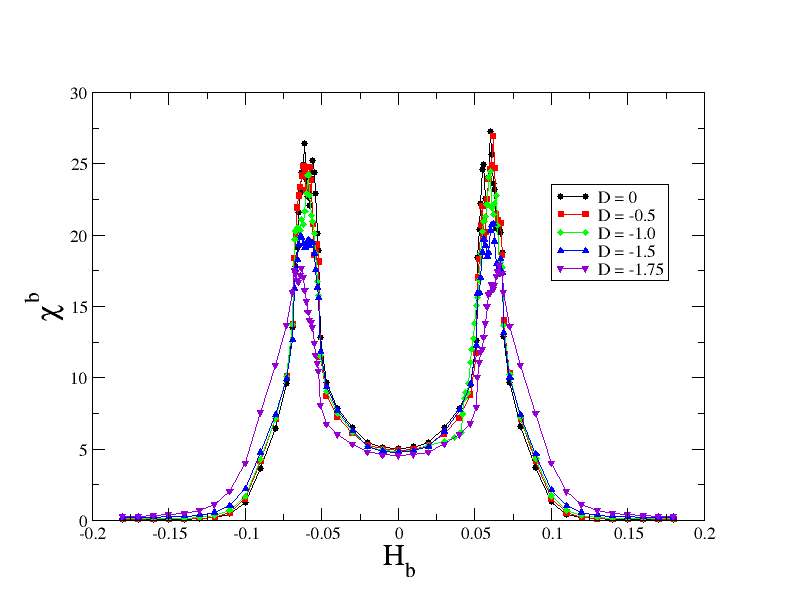}
    \caption{}
    \label{chibdistintosD}
\end{subfigure}
     \caption{Dependence of the sidebands on the crystal field $D$ 
     for $L = 128$. Here, $D \in[-1.75, 0]$, so zeros are significant. 
     For each value of $D$, we maintained the relationship $P \approx 4 P_c(D)$ and $T = 0.8 T_c(D)$. The values of $P_c(D)$ and $T_c(D)$ are given in Table \ref{tablapcritico}. (a) $\langle Q \rangle$ vs $H_b$. (b) $\chi^Q$ and (c) $\chi^b$ vs $H_b$. In both (b) and (c), 
     the peak heights decrease with increasing $D$, 
     while $|H^{peak}_b|$ increases slightly as $D$ approaches $D_t$. 
     }
        \label{distintosD}
\end{figure}

Additionally, by letting the crystal field $D$ approach its tricritical value 
from the positive side, we can study the behavior of the system as the density of zeros increases. Figure~\ref{distintosD} shows the results. The order parameter in Fig.~\ref{QdistintosD} exhibits an appreciable dependence on $D$ only in the regions near $|H^{peak}_b|$. 
The slope of $\langle Q \rangle$ around $|H^{peak}_b|$ decreases as $|D - D_t|$ 
is decreased. This is reflected in the decreasing heights of the scaled variance and susceptibility peaks in 
Figs.~\ref{chiQdistintosD} and \ref{chibdistintosD}, respectively. Our results also suggest that $|H^{peak}_b|$ increases as $D$ closely approaches $D_t$. This is shown by the data presented in 
Table~\ref{tablaD}, which contain the peak positions $|H^{peak}_b|$ for the values of $D$ analyzed. These results indicate that, as the density of zeros increases, larger bias fields are needed to generate the asymmetry responsible for
the sidebands, and even for these larger values of $H_b$, the heights of the fluctuation 
peaks are reduced. This is consistent with the assertion by Schick and Shih, that the 
effective interface tension should decrease in proportion to $|D - D_t|$ close to the
tricritical point. In fact, they calculated such a surface tension in a 
mean-field approximation \cite{schick1986}.   

\begin{figure}[H]
\centering
\begin{subfigure}{\columnwidth}
    \centering
    \includegraphics[width=0.55\columnwidth]{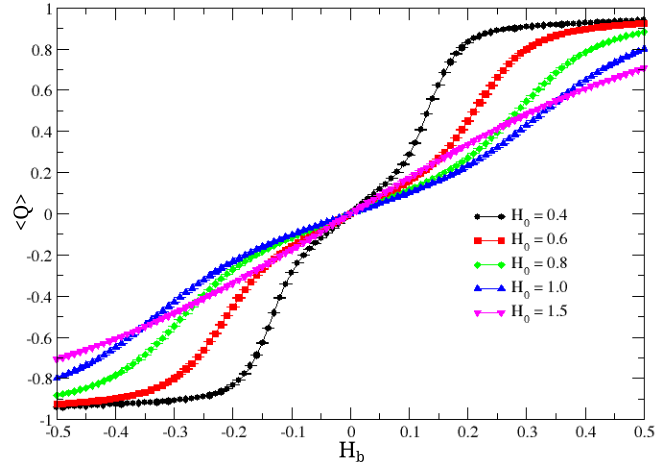}
    \caption{}
    \label{QdistintosH}
\end{subfigure}

\hspace{-50pt}
\begin{subfigure}{0.5\columnwidth}
    \centering
    \includegraphics[width=1.1\columnwidth]{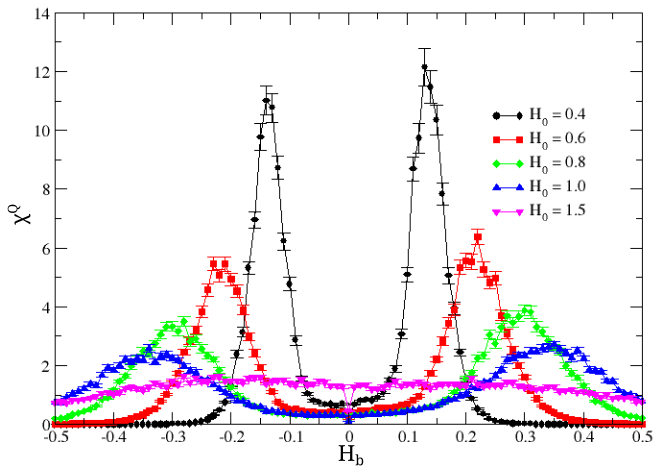}
    \caption{}
    \label{chiQdistintosH}
\end{subfigure}
\hspace{20pt}
\begin{subfigure}{0.5\columnwidth}
    \centering
    \includegraphics[width=1.1\columnwidth]{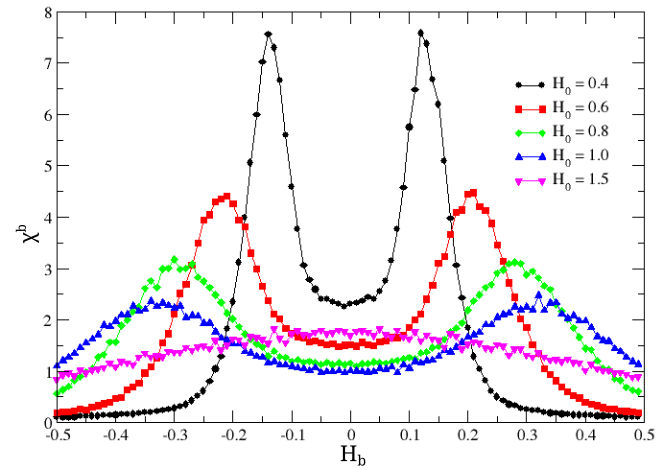}
    \caption{}
    \label{chibdistintosH}
\end{subfigure}
     \caption{Dependence of the sidebands on the amplitude $H_0$ of the oscillating magnetic field for $D = -1.5$, $L = 128$ and $P \approx 4 P_c(H_0)$ in each case. The critical periods for each value of $H_0$ are shown in Table \ref{tablaH0}. (a) $\langle Q \rangle$ vs $H_b$. (b) $\chi^Q$ and (c) $\chi^b$ vs $H_b$. In both (b) and (c), $|H^{peak}_b|$ increases with $H_0$, while 
     the peak heights decrease. For a sufficiently large $H_0$ with respect to $H_b$, the sideband peaks disappear and are replaced by a very low and broad central peak.}
        \label{distintosH}
\end{figure}

The sidebands also show a strong dependence on the amplitude $H_0$ of the oscillating field, as Fig.~\ref{distintosH} reflects. For each value of $H_0$, we maintained the relation $P(H_0) \approx 4P_c(H_0)$, where $P_c(H_0)$ can be found in Table \ref{tablaH0}. As $H_0$ increases, the derivatives of the curves for $\langle Q \rangle$ in Fig.~\ref{QdistintosH} decrease. 
$\chi^Q$ in 
Fig.~\ref{chiQdistintosH} and $\chi^b$ in Fig.~\ref{chibdistintosH} reflect this behavior. The peak heights become smaller, and $|H^{peak}_b|$ increases with $H_0$. As expected for larger $H_0$ values, larger bias fields are needed for the bands to appear. For a sufficiently large $H_0$ compared to $H_b$, the sideband peaks disappear and are replaced by a low, broad central peak. As Table \ref{tablaH0} suggests, higher values of $H_0$ result in smaller critical periods, and therefore, faster oscillating fields. As $H_0$ increases, the reduction of the critical period and the need for stronger bias together result in the eventual disappearance of the bands. These findings are in agreement with those obtained by Riego et al. \cite{Riego2018} for the kinetic Ising model.

\begin{table}[H]
    \centering
    \caption{Peak position $|H^{peak}_b|$ for different values of the crystal field $D$ in the kinetic BC model, corresponding to the second-order region in the phase diagram of the model at equilibrium. The peak positions are estimated from the data for $\chi^Q$, shown in 
    Fig.~\ref{chiQdistintosD}. $P\approx 4 P_c(D)$ and $T = 0.8T_c(D)$ in each case. 
    The values of the critical periods and temperatures for each value of $D$ are taken from Table~\ref{tablapcritico}.}
    \begin{tabular}{|c|c|}
       \hline
       $D$  & $|H^{peak}_b| (P \approx 4 P_c)$ \\
       \hline
       0 & 0.060 $\pm$ 0.001\\
       \hline
       $-$0.5 & 0.061 $\pm$ 0.002 \\
       \hline
       $-$1.0 & 0.062 $\pm$ 0.002\\
       \hline
       $-$1.5 & 0.063 $\pm$ 0.002\\
       \hline
       $-$1.75 & 0.067 $\pm$ 0.002\\
       \hline
    \end{tabular}
    \label{tablaD}
\end{table}

\begin{table}[H]
    \centering
    \caption{Critical periods $P_c$ for different values of the amplitude $H_0$ of the external oscillating field, and the position $|H^{peak}_b|$ of the fluctuation peaks for $P(H_0) \approx 4 P_c(H_0)$, and $D = - 1.5$. The peak positions are estimated from the data for $\chi^Q$, 
    shown in Fig.~\ref{chiQdistintosH}.} 
    \begin{tabular}{|c|c|c|}
       \hline
       $H_0$  & $P_c(H_0)$ & $|H^{peak}_b| (P(H_0) \approx 4 P_c(H_0))$ \\
       \hline
       0.2 & 205 $\pm 5$ & 0.063 $\pm$ 0.002\\
       \hline
       0.4  & 44 $\pm$ 5 & 0.14 $\pm$ 0.01 \\
       \hline
       0.6 &  23 $\pm$ 2 & 0.22 $\pm$ 0.01\\
       \hline
       0.8 & 15 $\pm$ 1 & 0.29 $\pm$ 0.01\\
       \hline
       1.0 & 10 $\pm 1$ & 0.35 $\pm$ 0.01\\
       \hline
       1.5 & 4 $\pm 1$ & - \\
       \hline
    \end{tabular}
    \label{tablaH0}
\end{table}

\subsection{Nucleation}
\label{nucleation}

\subsubsection{Magnetization reversal and decay mechanisms}
Nucleation theory describes the decay from a metastable to a stable phase through the formation of droplets of the stable phase. For ferromagnetic systems, spins with the same orientation as the external field, $(H(t)+H_b)$, form the stable phase, while spins with the opposite orientation form the metastable phase.

There are two main mechanisms for metastable decay \cite{Rikvold1994, Tomita1992, Richards1995}. Multi-droplet (MD) decay occurs by the independent nucleation and growth 
of many stable-phase droplets that eventually coalesce and span the lattice. On the other hand, if the decay occurs due to the nucleation and growth of a single droplet, the mechanism is known as single-droplet (SD). Since the spin reversal proceeds through a large number of independent processes, the MD mechanism is  
almost deterministic, and it is well described by the Kolmogorov-Johnson-Mehl-Avrami (\cite{Kolmogorov1937}, \cite{Johnson1939}, \cite{Avrami1_1939, *Avrami2_1939, *Avrami3_1939}) or KJMA theory. It dominates for large lattices and/or strong external fields,  
which cause the mean separation between droplets to be smaller than $L$ \cite{Richards1995}. In contrast, the SD mechanism is a stochastic process that dominates for small lattices and/or weak external fields, which cause the mean separation between droplets to be larger than $L$  \cite{Rikvold1994}. Buendía and Rikvold \cite{Buendia2017} showed that nucleation theory can be applied to study and explain the sidebands for the $S = 1/2$ Ising model. Since so far the BC model seems to behave very similarly to the Ising model, it makes sense to consider if nucleation theory also explains the appearance of the magnetic fluctuations in this model, and how the density of zeros might affect the process.

\begin{figure}[H]
\centering
\hspace{-40pt}
\begin{subfigure}{0.5\columnwidth}
    \includegraphics[width=1.1\columnwidth]{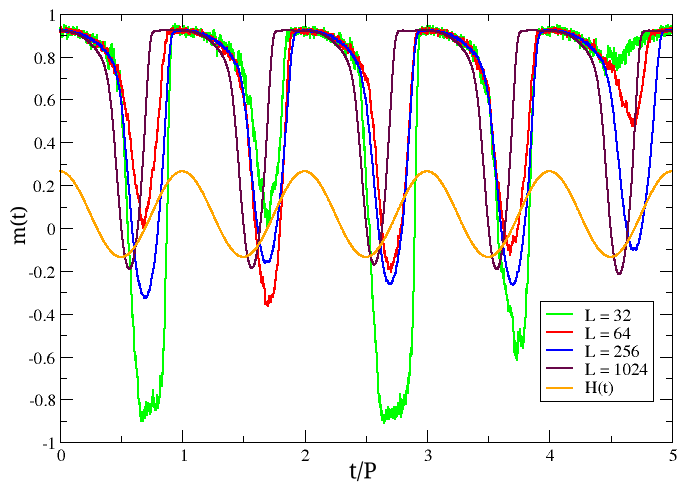}
    \caption{}
    \label{strong}
\end{subfigure}
\vspace{-7pt}
\hspace{20pt}
\begin{subfigure}{0.5\columnwidth}
    \includegraphics[width=1.1\columnwidth]{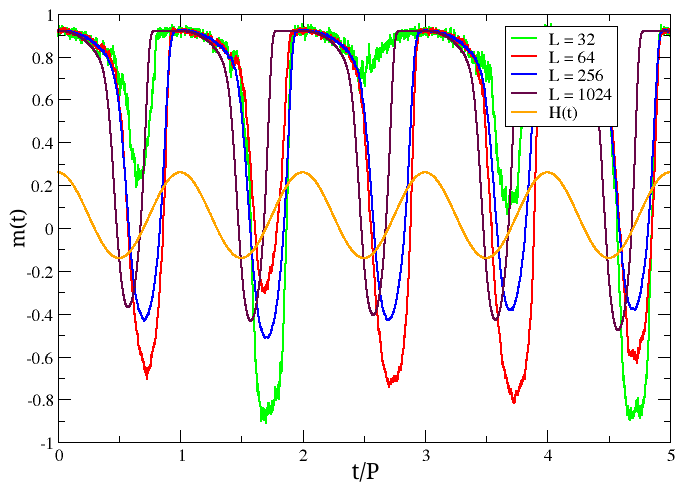}
    \caption{}
    \label{peak}
\end{subfigure}
\begin{subfigure}{0.5\columnwidth}
    \includegraphics[width=1.1\columnwidth]{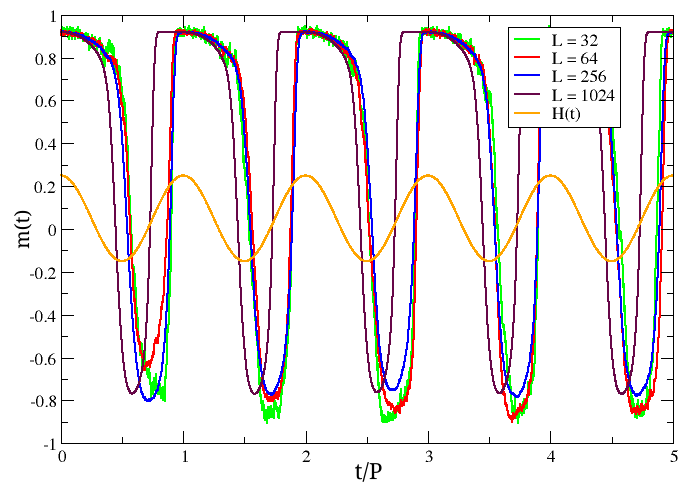}
    \caption{}
    \label{weak}
\end{subfigure}
     \caption{Time-dependent magnetization $m(t)$ for five cycles and different $L$ for 
     $D = -1.5$, $P = 820 \approx 4 P_c$, $H_0 = 0.2$ and different values of $H_b$. 
 Three different scenarios are examined: (a) $H_b = 0.065$, a strong bias, just above $|H^{peak}_b|$. (b) $H_b = 0.063$ near $|H^{peak}_b|$. (c) $H_b = 0.05$, a weak bias, below $|H^{peak}_b|$. The orange curve illustrates the total applied magnetic field $H(t) + H_b$. For an analysis of this figure, see Sec. \ref{nucleation}.}
        \label{m(t)nucleation}
\end{figure}

Figure \ref{m(t)nucleation} shows results of our simulation for the study of spin inversion in the kinetic BC model for $D = -1.5$, in the region $P > P_c$ $(P \approx 4 P_c)$, and its dependence on system size and bias. When the external field, $H(t) + H_b > 0$, the down-spin phase is metastable,  while when $H(t) +H _b < 0$, the up-spin phase is the metastable one. Since we select $H_b > 0$, it is more difficult for the spins to switch to the stable down-spin phase, creating an asymmetry in the decay times of the metastable phases. We call the state with an opposite sign to the bias the {\it disfavored} state. 

Figure \ref{strong} shows the results for $H_b$ above the maximum of the bands ($H_b > |H^{peak}_b|$). For the smallest lattices, $L = 32$ and $L = 64$, the magnetization exhibits a stochastic behavior, typical of the SD regime, and the maximum negative value of $m(t)$ varies considerably from cycle to cycle. Occasionally, for $L=32$ a large number of spins switch toward the disfavored state, and the maximum negative value of $m(t)$ may become as large as $-0.9$.
For larger lattices, $L \ge 128$, the magnetization reversal toward the unfavorable state is more deterministic. Fewer the spins switch toward the disfavored phase, so the magnetization does not reach large negative values. In these cases, the maximum negative value of the magnetization approaches $-0.2$ on average.  

For $H_b$ near $|H^{peak}_b|$, Fig.~\ref{peak}, the stochastic SD behavior again prevails for the smallest size, $L = 32$, but the deterministic MD behavior starts to appear for systems as small as $L = 64$. In this case, more spins switch toward the disfavored phase, resulting in larger negative values of $m(t)$, such that for the larger systems ($L \geq 256$), the maximum negative value of $m(t)$ remains close to $-0.4$ on average. However, even for the largest system, $L = 1024$, 
the fluctuations from cycle to cycle are still significant and support the conclusion that the 
peaks are a manifestation of stochastic resonance. 

For a very weak value of $H_b$ ($H_b < |H^{peak}_b|$), Fig.~\ref{weak}, the size dependence becomes less important. Even the smallest system, $L = 32$, behaves almost
deterministically. All the lattices switch to the disfavored phase. As in the previous cases, the maximum negative value of $m(t)$ becomes smaller for the larger sizes, but the fluctuations are reduced. For $L = 1024$, this value is approximately $-0.77$.

These results are consistent with the principles of nucleation theory. For small $L$, the stochastic behavior, characteristic of a SD mechanism, dominates, while for large $L$, the deterministic MD mechanism prevails. Additionally, our results suggest that, for lattices of intermediate size, there is a crossover from the SD to the MD nucleation mechanism as $H_b$ decreases. 

\begin{figure}[H]
\centering
\hspace{-40pt}
\begin{subfigure}{0.5\columnwidth}
    \includegraphics[width=1.1\columnwidth]{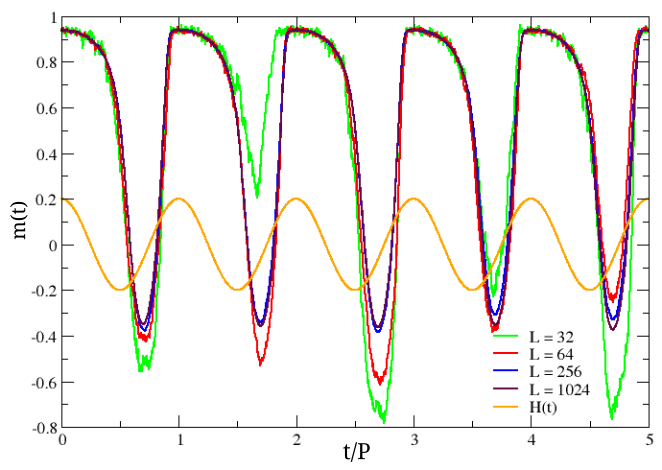}
    \caption{}
    \label{strong1.75}
\end{subfigure}
\vspace{-7pt}
\hspace{20pt}
\begin{subfigure}{0.5\columnwidth}
    \includegraphics[width=1.1\columnwidth]{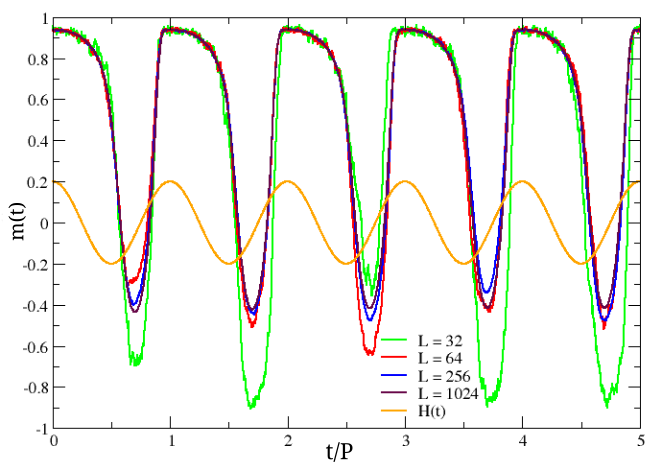}
    \caption{}
    \label{peak1.75}
\end{subfigure}
\begin{subfigure}{0.5\columnwidth}
    \includegraphics[width=1.1\columnwidth]{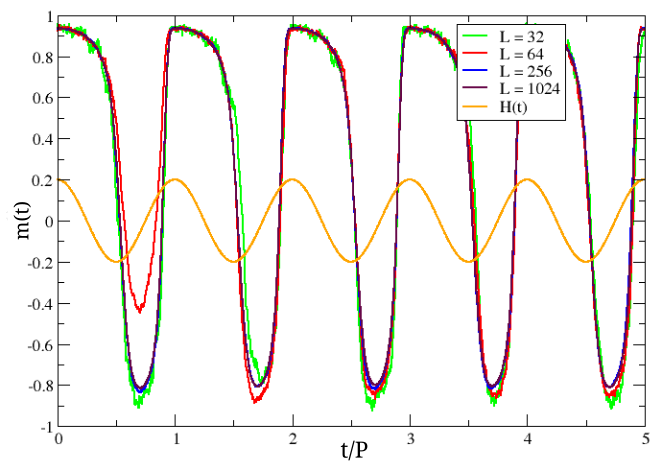}
    \caption{}
    \label{weak1.75}
\end{subfigure}
     \caption{Time-dependent magnetization $m(t)$ for five cycles and different lattice sizes for $D = -1.75$, $P = 460 \approx 4 P_c$, $H_0 = 0.2$ and different values of $H_b$. For this period length, the peaks occur at $|H^{peak}_b| \approx 0.067$. (a) $H_b = 0.068$, a strong bias, just above $|H^{peak}_b|$. (b) $H_b = 0.066$ near $|H^{peak}_b|$. (c) $H_b = 0.052$, a weak bias, below $|H^{peak}_b|$. There are no qualitative differences between these results and those depicted in Fig.~\ref{m(t)nucleation} for $D = -1.5$.}
        \label{m(t)nucleationotroD}
\end{figure}

Figure \ref{m(t)nucleationotroD} shows the same plots as Fig.~\ref{m(t)nucleation}, now for $D = -1.75$, in which case the density of zeros is greater than for $D = -1.5$. 
We study the effect of changing $D$ by comparing Fig.~\ref{m(t)nucleation} and 
Fig.~\ref{m(t)nucleationotroD}, and for these values of $D$, there are no qualitative 
differences in the behavior of the system. For $D = -1.75$, 
the fluctuation peaks, and therefore the 
sidebands, occur at larger values of the bias. 
This behavior is expected since, as the density of zeros increases, there are more spins for which it is easier to follow the external field.

In the following, we will keep the value of $D = -1.5$.

\begin{figure}[H]
\centering
\begin{subfigure}{0.75\columnwidth}
    \centering
    \includegraphics[width=\columnwidth]{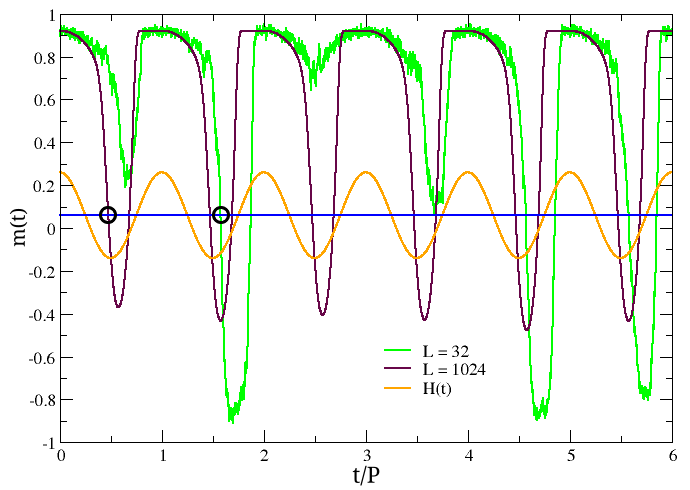}
    \caption{}
    \label{reference}
\end{subfigure}
\vspace{-15pt}
\begin{subfigure}{0.38\columnwidth}
    \centering
    \includegraphics[width=\columnwidth]{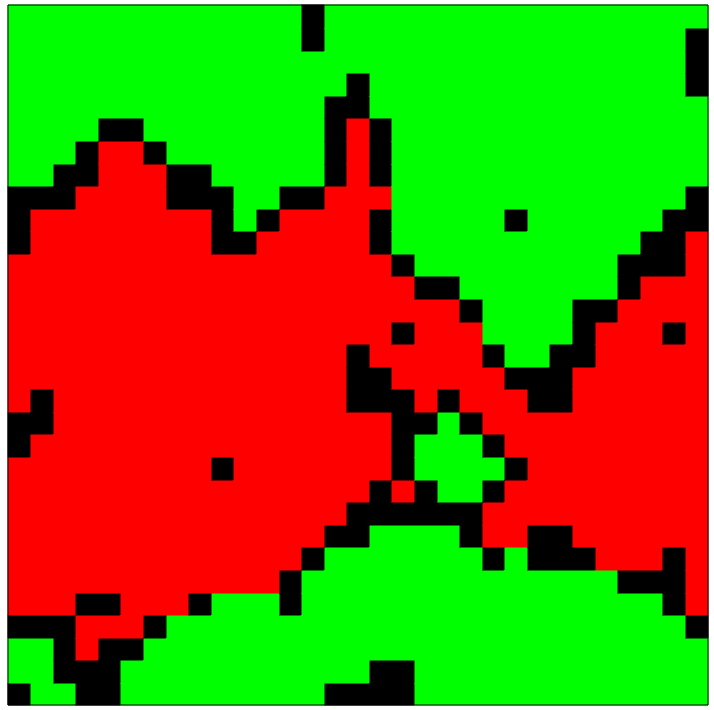}
    \caption{}
    \label{SDnucleation}
\end{subfigure}
\vspace{-5pt}
\hspace{30pt}
\begin{subfigure}{0.38\columnwidth}
    \centering
    \includegraphics[width=\columnwidth]{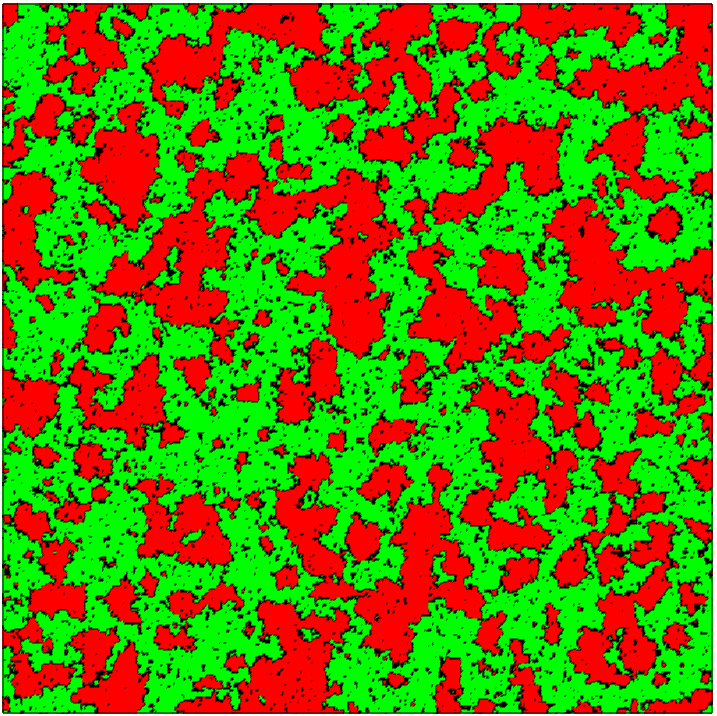}
    \caption{}
    \label{MDnucleation}
\end{subfigure}
     \caption{(a) $m(t)$ over six cycles for two different lattice sizes, $L = 32$ and $L = 1024$. The total applied field $H(t) + H_b$, with $D = -1.5$, $P = 820 \approx 4 P_c$ and $H_b = 0.063 \approx |H^{peak}_b|$, is represented in orange. Snapshots of the lattice were taken for both sizes when $m(t)$ first reached the threshold of 0.063 (indicated by the blue horizontal line). The times of capture are marked by black circles. For this threshold value, $H(t) + H_b < 0$, so the disfavored configuration of down-spins is the stable state, which is shown in red. Up-spins 
     are shown green, and zeros in black. (b) Snapshot for $L = 32$. The decay occurs by the nucleation and growth of a single droplet of the stable state (SD mechanism). 
     (c) Snapshot for $L = 1024$. The decay occurs by simultaneous nucleation and growth of multiple droplets of the down-spin phase (MD mechanism).}
        \label{snapshot}
\end{figure}

Figure \ref{snapshot} further illustrates the difference between the two nucleation mechanisms for $H_b \approx |H^{peak}_b|$. Snapshots of lattices of size $L = 32$ and $L = 1024$ were captured the first time $m(t)$ reached a threshold value of $m(t) = 0.063$, as indicated by the two circles in Fig.~\ref{reference}. For this value of $m(t)$, the total applied field is negative, and the snapshots display the growth of droplets of the down-spin phase, which is disfavored by the positive bias. This threshold value is chosen in a range 
to catch the droplets at a time when they grow 
rapidly, but their probability of percolating through the system is still relatively low.

For $L = 32$ in Fig.~\ref{SDnucleation}, we observe a single droplet of down-spins that expands throughout the entire system, signaling an SD regime. The time series in Fig.~\ref{reference} reflects the stochastic behavior associated with this decay mechanism. Of the six cycles plotted, we only observe a switching of the majority of spins in three of them. 

The system with $L = 1024$ looks quite different. Instead of a single droplet, the snapshot of 
Fig.~\ref{MDnucleation} shows the growth of multiple smaller clusters from the equilibrium state. As time passes, some droplets will grow and coalesce, while others will still grow independently as the spins reverse toward the stable state, in a way typical of the MD regime. The time series in Fig.~\ref{reference} shows the deterministic behavior of this regime for $L = 1024$. The magnetization reaches the threshold value in all cycles and, except for minor fluctuations, its maximum negative value is the same in each cycle. 

We notice, both for the SD and MD regimes, that the zero spins are preferentially located at the interface between the up and down spin clusters. This effect, which does not have an equivalent in the Ising model, suggests that the zero spins act as a surfactant, leading to a reduced effective surface tension \cite{schick1986}. 

\begin{figure}[H]
\centering
\hspace{-45pt}
\begin{subfigure}{0.4\columnwidth}
    \centering
    \includegraphics[scale=0.45]{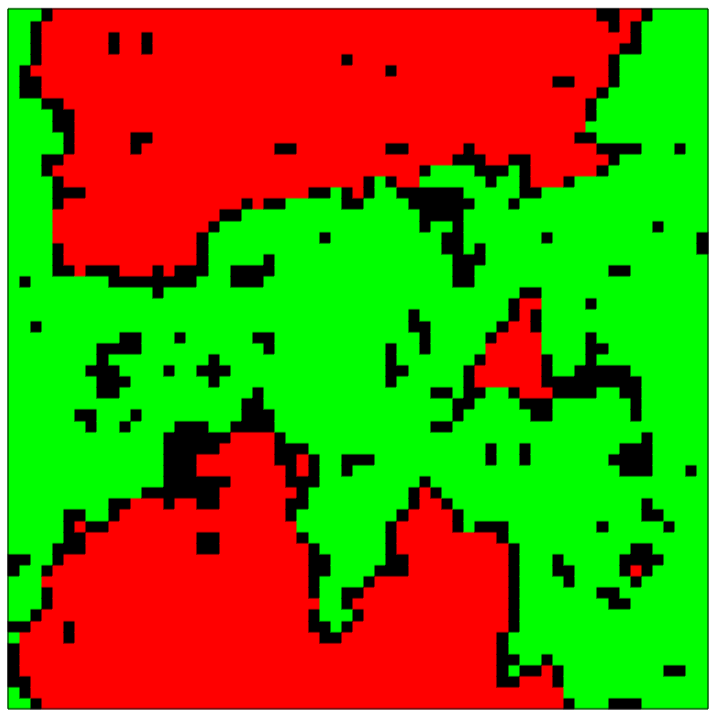}
    \caption{}
    \label{64strong}
\end{subfigure}
\hspace{45pt}
\begin{subfigure}{0.4\columnwidth}
    \centering
    \includegraphics[scale=0.45]{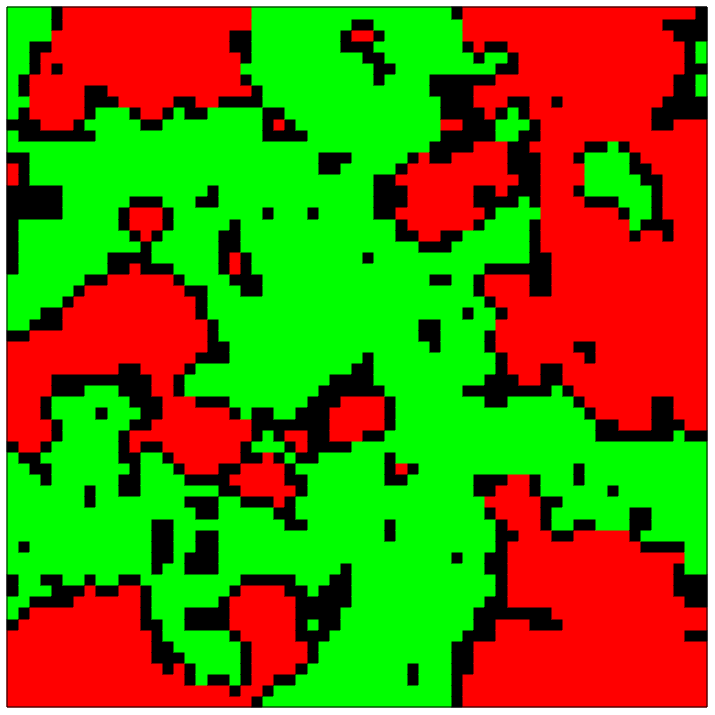}
    \caption{}
    \label{64weak}
\end{subfigure}
\caption{Nucleation of clusters of the stable configuration for different values of $H_b$ and $L = 64$, with $D = -1.5$ and $P = 820 \approx 4 P_c$. The images were captured the first time that $m(t)$ reached the threshold value of $m(t) = 0.063$, in which case $H(t) + H_b < 0$, and the stable configuration is the down-spin phase, which is shown in red. The up-spin phase is shown in green and zeros in black. (a) $H_b$ = 0.065, on the strong-bias side of the fluctuation peaks. (b) $H_b = 0.05$, on the weak-bias side of the fluctuation peaks. The images indicate a crossover between the SD regime for strong bias, and the MD regime for a weak bias.}  
\label{snapshots64}
\end{figure}

To study in more detail the dependence of the decay mechanism on $H_b$, 
Fig.~\ref{snapshots64} shows the nucleation of the down-spin phase for $L = 64$ and different values of $H_b$. As Fig.~\ref{64strong} illustrates, the single-droplet mechanism seems to dominate for a strong bias, just above the fluctuation peak. However, as $H_b$ decreases below the fluctuation peak, smaller clusters begin to emerge, indicating a transition toward a MD mechanism, as evidenced in Fig.~\ref{64weak}. These findings agree with the results previously discussed in 
Fig.~\ref{m(t)nucleation} on the magnetization reversal for different values of $H_b$. Notably, for intermediate lattice sizes, such as $L = 64$, the spin reversal process becomes more deterministic and consistent with a MD mechanism as $H_b$ decreases. 

\subsubsection{Dependence of $|H^{peak}_b|$ on $P$}

Our previous nucleation results can be explained quantitatively, studying the dependence of $|H^{peak}_b|$ on the period length of the external oscillating field. Buendía and Rikvold \cite{Buendia2017} employed nucleation theory on the kinetic Ising model subjected to a sinusoidal field and a
bias and, by comparing the magnetization switching rates for the two nucleation mechanisms near the maximum disfavored field, they found the following relation for very large periods:

\begin{equation}
     P \sim L^{-a} \exp{\left(\frac{1}{b} \frac{\Xi_0}{H_0 - |H^{peak}_b|} \right)}\quad ,
     \label{ecnucleation}
\end{equation}
with $a = 2$ and $b = 1$ in the SD regime, and $a = 0$ and $b = 3$ in the MD regime. $\Xi_0$ is the field-independent part of the free energy cost of a critical droplet divided by $k_B T$ \cite{Sides1999}. In two dimensions and for a temperature of $T = 0.8T_c$, $\Xi_0 \approx 0.506$

This relation suggests a linear relationship between $\sfrac{1}{(H_0 - |H^{peak}_b|)}$ and 
$\ln{P}$ for large periods, such that the ratio between the slopes for the MD and SD regime is 3. This linear trend was confirmed for the kinetic Ising model \cite{Buendia2017} for periods between $P = 14000$ and $P = 28000$ and the ratio of the slopes was found to be approximately 2.867.

Likewise, a recent effective field study \cite{Yuksel2022} on the kinetic BC model in the presence of a sinusoidally oscillating magnetic field and a constant bias, also found the linear relationship to hold for long periods and obtained a ratio of 2.236.

\begin{figure}[H]
\centering
\hspace{-35pt}
\begin{subfigure}{0.45\textwidth}
    \centering
    \includegraphics[scale=0.45]{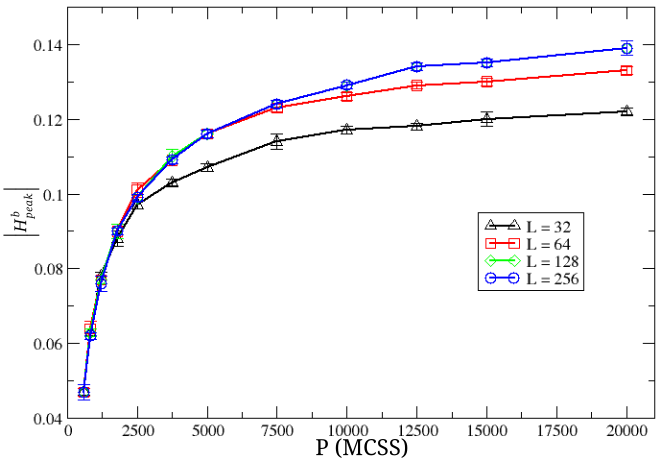}
    \caption{}
    \label{peakslog}
\end{subfigure}
\hspace{17pt}
\begin{subfigure}{0.45\textwidth}
    \centering
    \includegraphics[scale=0.45]{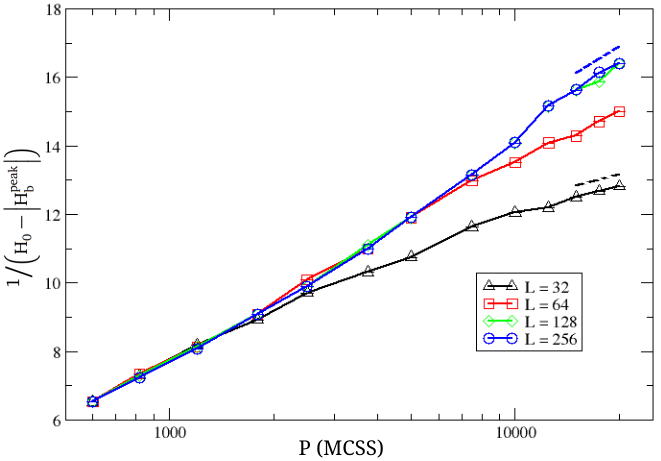}
    \caption{}
    \label{lognucleation}
\end{subfigure}
\caption{Dependence of the fluctuation peaks position on the period $P$ ($P > P_c \approx 205$) for different lattice sizes, $D = -1.5$ and $H_0 = 0.2$. (a) $|H^{peak}_b|$ vs $P$. (b) $\sfrac{1}{(H_0 - |H^{peak}_b|)}$ vs $\log P$. For large periods ($P$ between 15000 and 20000), we obtained a linear trend, as expected according to Eq.~(\ref{ecnucleation}). The blue and black dashed lines represent the slopes of the curves for the MD regime ($L = 256$) and the SD regime ($L = 32$), respectively. The ratio between the slopes is approximately 2.406.}  
\label{nucleationpeaks}
\end{figure}

The results of our simulations for the peak positions $|H^{peak}_b|$ of $\chi^Q$ vs $P$ 
are plotted in 
Fig.~\ref{peakslog} for different lattice sizes. For small periods, $|H^{peak}_b|$ increases rapidly, while for large periods, there is little variation in the peak position. For the smaller lattices, and particularly for large periods ($P > 2500$), finite-size effects are apparent. 
These effects become negligible for larger lattices, where the MD switching mechanism dominates. 

Figure \ref{lognucleation} shows our results for $\sfrac{1}{(H_0 - |H^{peak}_b|)}$ vs $\log{P}$. As predicted by Eq.~(\ref{ecnucleation}), for large $P$ (between $P = 15000$ and $P = 20000$), we obtained a linear trend. The dotted lines represent the slopes of the curves for $L = 32$ and $L = 256$ (black and blue, respectively). The ratio between the slopes for the MD and SD regimes is approximately 2.406. This ratio is lower than the one found by Buendía and Rikvold \cite{Buendia2017} for the kinetic Ising model, but higher than that reported by Yüksel et al. \cite{Yuksel2022} for the kinetic BC model under effective field theory. Given that zeros influence the position of the peaks, it is expected that the ratio between the slopes only approximately equals the predicted value of 3 according to Eq.~(\ref{ecnucleation}). 
Additionally, it is worth noting that Fig.~\ref{nucleationpeaks} is very similar to that obtained by previous studies on the kinetic Ising model \cite{Buendia2017, Rikvold1994}.\\

Based on the results of this section, we can therefore conclude that, as in the Ising model, the sidebands are not due to a critical phenomenon, but rather to a change in the nucleation mechanism with the bias. Furthermore, the decrease of the ratio between the slopes in 
Fig.~\ref{lognucleation} compared to the results for the kinetic Ising model \cite{Buendia2017}, indicates that the zeros affect the nucleation and growth mechanisms.

\section{Summary and Conclusion}
\label{conclusion}
Although it is well known that the kinetic Ising model belongs to the same universality class as the equilibrium Ising model \cite{Sides1999, PhysRevLett.81.834, PhysRevE.63.016120}, where the period of the time-oscillating field of the kinetic model plays the same role as the temperature in the equilibrium model, recent experimental studies on thin uniaxial Co films \cite{Riego2017} show a fascinating behavior of the kinetic Ising model in the presence of a constant bias field, which has no equivalent in the equilibrium counterpart. When a constant field, $H_b$, is added to a slow oscillating field of period $P$ above the critical value $P_c$, the kinetic model manifests metamagnetic anomalies that appear in the form of two large peaks in the susceptibilities, symmetric around $H_b = 0$. The findings of the experimental study conducted by Riego et al. \cite{Riego2017} clearly indicated that these anomalies are not a critical phenomenon. Additional numerical studies \cite{Buendia2017} demonstrated that the fluctuations are not due to residual magnetization of the experimental system, but a stochastic resonance phenomenon associated with a change of the magnetization reversal mechanism induced by the asymmetry introduced by the bias field. Since these findings can have important effects in the performance of devices based on magnetization reversal induced by time-oscillating fields, we wondered if this phenomenon also appears in
more complex Ising-type models such as the kinetic Blume-Capel (BC) model, which also manifests a dynamic phase transition that belongs to the Ising universality class \cite{Vatansever2018, PhysRevE.97.062146, YUKSEL2020167249, PhysRevE.104.024108}.

In a recent mean-field study \cite{Yuksel2022}, Y{\"u}ksel et al.\ found that a kinetic BC model in the absence of crystal field (i.e.\ $D = 0$) also exhibits 
symmetric fluctuation peaks when a constant bias field is added. We therefore decided to 
perform a nonperturbative study of this model, following the steps
of the detailed Monte Carlo study of the kinetic Ising model by Buendía and Rikvold \cite{Buendia2017}. We simulated the kinetic BC model in two dimensions, with a 
crystal field $D$ between 0 and $-1.75$ and ferromagnetic interactions, in the presence of an oscillating magnetic field and a constant bias.

For $P < P_c$, the kinetic and equilibrium BC models exhibit analogous behavior in terms of the magnetic response to changes in the respective conjugate fields. For the disordered region ($P > P_c$) and $D_t < D \leq 0$, which is our focus in the present study, the susceptibility 
curves display two sideband peaks, symmetrical with respect to zero bias, in agreement with the findings of previous studies on the kinetic Ising model \cite{Riego2017, Buendia2017} and the recent work on the BC model by Yüksel et al. \cite{Yuksel2022}. 
As in the Ising model \cite{Riego2018}, the sidebands are strong for weak oscillating fields, but they disappear for very large field amplitudes or, equivalently, for rapidly oscillating fields. 

We did not observe any indications, such as hysteretic behavior or divergent finite-size effects, that would suggest that these bands are the result of a phase transition. Rather, we applied nucleation theory to demonstrate that these fluctuations are associated with a change in the magnetization switching mechanism between the stable and metastable states. 
As $|H_b|$ is increased, the average separation between the 
growing droplets becomes larger than $L$, and the switching mechanism changes from the deterministic MD mechanism to the stochastic SD mechanism.  

Our results suggest a dependence of the fluctuation peaks of the susceptibilities on the crystal field. As the density of zeros increases ($D$ approaching $D_t$ from the positive side), 
the size of the fluctuations is radically reduced. Although less clear, it also appears that larger bias fields are necessary for the 
peaks to appear. (See Fig.~\ref{chiQdistintosD}.) 

Furthermore, as $D$ approaches its 
tricritical value, the metastable lifetime measured in field-reversal simulations below $T_c$ 
has been observed to decrease \cite{Vatansever2018}. 
As shown in Figs.~\ref{snapshot} and \ref{snapshots64} in our present work, 
the zeros concentrate on the interface 
between the regions of the up-spin and the down-spin phases. 
The same effect at very low temperatures has been observed as 
a ``picture-frame'' configuration of critical droplets by Cirillo et al.\ \cite{CIRI96,CIRI24} 
and in a 2D mean-field model of microemulsions by Schick and Shih \cite{schick1986}. 
These observations together give strong support to our assumption that the zero phase acts as 
a surfactant to provide a reduced effective surface tension between the up-spin and 
down-spin phases. 

Our study provides further evidence that the equivalence between DPT and TPT is only valid near the critical point. Outside this region, kinetic models can present a much richer behavior that does not have a parallel in the equilibrium ones. It also demonstrates that metamagnetic anomalies are not a phenomenon exclusive to the $S=1/2$ Ising model, but rather a general property of DPTs. We hope that these results will contribute to a better understanding of dynamic phase transitions and their similarities and differences with thermodynamic transitions. 

\section*{Acknowledgments}
G.M.B. expresses her appreciation for support and hospitality
at the PoreLab and NJORD Centres of the Department of Physics at the University of Oslo.

Work at the University of Oslo was supported by the Research Council of Norway through the Center of Excellence funding scheme, Project No. 262644.


%

\newpage

\end{document}